\documentclass[iop]{emulateapj}
\usepackage{comment}
\usepackage{amsmath}
\usepackage{amssymb}
\usepackage{epstopdf}
\usepackage{hyperref}
\usepackage{natbib}
\usepackage{verbatim}
\usepackage{tablefootnote}
\usepackage{graphicx}

\newcommand{\icarus}{Icarus} 

\slugcomment{Received 2015 May 20; accepted 2015 June 12; }

\shorttitle{ A method to measure exoplanet spins }
\shortauthors{ Nikolov et al. }

\begin{document}


\title{ Radial velocity eclipse mapping of exoplanets }


\author{ Nikolay~Nikolov and Felix~Sainsbury-Martinez}
\affil{Astrophysics Group, School of Physics, University of Exeter, Stocker Road, Exeter EX4 4QL, UK}
\email{ nikolay@astro.ex.ac.uk  }









\begin{abstract}
Planetary rotation rates and obliquities provide information regarding the history of planet formation, but have not yet been measured for evolved extrasolar planets.\,Here we investigate the theoretical and observational perspective of the Rossiter-McLauglin effect during secondary eclipse (\textit{RMse}) ingress and egress for transiting exoplanets. Near secondary eclipse, when the planet passes behind the parent star, the star sequentially obscures light from the approaching and receding parts of the rotating planetary surface. The temporal block of light emerging from the approaching (blue-shifted) or receding (red- shifted) parts of the planet causes a temporal distortion in the planet's spectral line profiles resulting in an anomaly in the planet's radial velocity curve.\,We demonstrate that the shape and the ratio of the ingress-to-egress radial velocity amplitudes depends on the planetary rotational rate, axial tilt and impact factor (i.e. sky-projected planet spin-orbital alignment). In addition, line asymmetries originating from different layers in the atmosphere of the planet could provide information regarding zonal atmospheric winds and constraints on the hot spot shape for giant irradiated exoplanets. The effect is expected to be most-pronounced at near-infrared wavelengths, where the planet-to-star contrasts are large. We create synthetic near-infrared, high-dispersion spectroscopic data and demonstrate how the sky-projected spin axis orientation and equatorial velocity of the planet can be estimated. We conclude that the \textit{RMse} effect could be a powerful method to measure exoplanet spins. 
\end{abstract}






%
%
%
%
%
\section{Introduction}\label{sec:introsec}
%
%
%



Planetary rotation rates and axial tilts are critical parameters in determining seasonal climate variations \citep{williams97, cowan12} and are relevant to the planet formation and evolution history \citep{agnor99}. By definition the planet rotational rate and axial tilt (e.g. obliquity) are respectively the time it takes for a complete revolution and the angle between the planetary spin angular momentum and the planet's orbital angular momentum. Planets with obliquity $<90^{\circ}$ and $>90^{\circ}$ are considered to have prograde and retrograde rotation, respectively. Planet spins are well-constrained for the Solar system planets, spanning a wide range of rotational rates and axial tilts \citep{cox2000} and are considered to reflect the unique formation and evolutionary history of each planet \citep{laskar1993}. Planets accumulate rotational angular momentum from the relative motions of accreted material. The stochastic nature of planetary accretion from planetesimals allows for a random component to the net spin angular momentum. Prograde angular momentum (i.e. spin) could be accumulated by a planet on a circular orbit within a uniform surface density disk of small planetesimals. The planet clears a gap and thus accretes a larger fraction of material from the edges of its accretion zone. Retrograde spins are considered to originate from giant impacts during the early stages of planet formation. Therefore constraints on the planet spins are of high scientific interest, because deviations of the axes of rotation may have been caused by impacts of large bodies during their early history. Even the axes of rotation of the gaseous planets may have been affected by impacts on their rocky cores before these planets accumulated their large atmospheres of hydrogen and helium. Such atmospheres are considered to be accreted hydrodynamically, in flows quite different from those which govern the dynamics of planetesimals and lead to prograde rotation \citep{dePater2001, Faure_2007}.

Exoplanet tidal theory predicts obliquity erosion, on time scales $<1$\,Gyr, for planets on $\lesssim10$\,day orbits around normal low-mass stars, i.e. preventing seasonal variations \citep{heller11a, heller11b}. Observational methods have been proposed to probe the rotation rates and obliquities for exoplanets from oblateness measurements and variability due to surface inhomogeneity e.g. \cite{hui02, seager02, Barnes03, palle08, kawahara10, fuji12}, but require precisions in excess of $\sim0.1$\,$\mu$mag. 

Currently a rotational rate has been probed for only one extrasolar planet - the young fast rotator $\beta$\,Pic\,b, \citep{snellen14} while axial tilts have not yet been measured. Ground-based high-dispersion spectroscopy ($\text{R}\ge20,000$) in the near-infrared has recently become successful in characterising the atmospheres of hot Jupiters \citep{Snellen10, Brogi12, Birkby13, Brogi14, Schwarz15}. At high spectral resolution molecular absorption bands are resolved into individual lines allowing their robust identification by line matching with model templates. As the planet orbits its star, the radial component of the planet orbital velocity changes by tens of km\,s$^{-1}$, enabling a discrimination of the Doppler shifted planet spectrum from the steady telluric contamination. The planet signal is then extracted by cross-correlating the data with model spectra obtained by mixing the expected spectroscopically active trace gases in hot-Jupiter atmospheres and assuming a range of vertical temperature pressure profiles.

\cite{kawahara12} considered the effect of planetary spin on the planetary radial velocity in dayside spectra of exoplanets, simulated the effect and concluded that planetary radial velocity could be a powerful means for constraining planet spins. In this paper we describe the potential of the Rossiter-McLauglin effect during secondary eclipse (\textit{RMse}) combined with near-infrared high-dispersion spectroscopy to provide constraints on exoplanet rotation and obliquity. 

The rest of the paper is organised as follows. Section~\ref{sec:RM} describes the rotational effect in transiting exoplanets.  In Section~\ref{sec:RM_F} we derive the planet radial velocity anomaly due to the \textit{RMse} effect. We choose the methodology of \cite{ohta05} who found an analytic solution for the RM effect at primary transits and show that their solution can equivalently be applied to the problem discussed here. Section~\ref{sec:reas} details the amplitude and shape of the \textit{RMse} effect and discusses potential targets. Section~\ref{sec:obs_persp} discusses the potential application of the method in light of the available and future instrumentation. Finally Section~\ref{sec:discussionsec} is devoted to our conclusions. \\


%
%
%
%
%
%
%
%
\section{RM effect at planet secondary eclipse}\label{sec:RM}
In its nature, the RM effect arises because of stellar rotation \citep{rossiter1924, mcLaughlin1924}. During a stellar occultation or a planetary transit, portions of the rotating star surface are temporarily obscured, causing the removal of particular radial velocity components from the stellar broadened absorption lines leading to a temporal radial velocity anomaly \citep{winn05}. The same effect is expected to take place with a rotating planet at secondary eclipse. When the planet passes behind the parent star the light from the approaching and receding parts of the rotating planetary surface sequentially enter/exit the geometric shadow of the star. The temporal block of light emerging from the approaching (blue-shifted) or receding (red-shifted) parts of the planet causes temporal distortion in the line profiles of the planet's spectrum leading to an anomaly in the planet radial velocity curve.\,The shape and amplitude of this anomaly depend on the planet rotation rate, axial tilt and spatial orientation (i.e. planet spin-orbital alignment). Measurements of rotational rates and axial tilts are of high scientific interest as they provide constraints on exoplanet formation and evolution.\\ 

%
%
%
%
%
%
%
%
%
%
%
%
%
%
%
\begin{deluxetable}{lll} 
\tablecolumns{3} 
\tablecaption{ \textsc{List of Notation} \label{table:results1}} 
\tablehead{ \colhead{Variables}  & \colhead{Definition} & \colhead{Meaning} }
%
\multicolumn{3}{c}{ Orbital Parameters }\\[2.5pt]
\hline\\[-7pt]
$m_s\,..........$                          &    \S\,\ref{sec:RM_F}     &   Star mass\\
$m_p\,..........$                         &    \S\,\ref{sec:RM_F}     &   Planet mass\\
$P\,............$                           & \S\,\ref{sec:RM_F}               &            Orbital period \\
$a\,.............$                           &    Fig.\,\ref{fig:FIG1}$^\ast$     &   Semimajor axis\\
$e\,.............$                           & Fig.\,\ref{fig:FIG1}$^\ast$        & Planet orbital eccentricity\\
$\overline{\omega}\,............$  & Fig.\,\ref{fig:FIG1}$^\ast$     &   Argument of periastron\\
$E\,............$                           &   Fig.\,\ref{fig:FIG1}$^\ast$    &      Eccentric anomaly \\
$n\,.............$                          & Fig.\,\ref{fig:FIG1}$^\ast$ & Mean motion \\
$M\,...........$                           & Fig.\,\ref{fig:FIG1}$^\ast$ & Mean anomaly \\
$\tau\,.............$                      & Fig.\,\ref{fig:FIG1}$^\ast$ & time of pericentre passage \\
$i\,.............$                            & Fig.\,\ref{fig:FIG1}\,a      &      orbital inclination\\
$f\,............$                            & Fig.\,\ref{fig:FIG1}\,a    & True anomaly\\
$r_p\,...........$                         & Fig.\,\ref{fig:FIG1}\,a     &      planet to star distance\\
\cutinhead{Internal Parameters of Star and Planet}
$I_p...........$      &   Fig.\,\ref{fig:FIG1}\,a    &     Planet spin-to-$y$-axis angle \\
$\Omega_p..........$       & Fig.\,\ref{fig:FIG1}\,a & Planet angular velocity\\
$\lambda_p..........$      & Fig.\,\ref{fig:FIG1}\,b  & Sky-projected spin-orbit angle \\
$R_s..........$      & \S\,\ref{sec:RM_F}  & Stellar radius\\
$R_p..........$      & \S\,\ref{sec:RM_F}  & Planet radius\\
$V_p..........$      & \S\,\ref{sec:RM_F}  & Planet surface velocity, $R_p \Omega_p$ \\
\cutinhead{Mathematical notation}
$\hat{n}_p\,.........$ &  Fig.\,\ref{fig:FIG1}\,a,\,b  & Normal vector to the planet orbit\\
x$_s.........$ &  \S\,\ref{sec:RM_F}  & Position of the star\\
x$_p.........$ &  \S\,\ref{sec:RM_F}  & Position of the planet\\
$\gamma............$ &  \S\,\ref{sec:RM_F}  & Star to planet ratio $R_{\ast}/R_p$\\
$\eta_s..........$ & Eq\,\ref{eq:eq8} & See Fig.\,\ref{fig:FIG1}\,c\\
$x_0..........$ & Eq\,\ref{eq:eq13} & See Fig.\,\ref{fig:FIG1}\,c\\
$z_0..........$ & Eq\,\ref{eq:eq14} & See Fig.\,\ref{fig:FIG1}\,c\\
$\zeta_s..........$ & Eq\,\ref{eq:eq15} & See Fig.\,\ref{fig:FIG1}\,c\\
\tablecomments{see Fig.\,1 in \cite{ohta05} for all quantities marked with ($^\ast$) and are defined as expected.} 
\label{table:notation}
\end{deluxetable}
%
%
\section{Formalism of the effect}\label{sec:RM_F}
To describe quantitatively the radial velocity anomaly caused by the \textit{RMse} effect we assume a two-body problem with a central star and a planet of masses $m_{s}$ and $m_{p}$, respectively. We refer the reader to Fig.\,1 in \cite{ohta05} for a schematic illustration of the top view of the planetary orbit and their equations (1) to (7). The orbital velocity of the planet as a function of time with respect to the star, up to $O(e)$ as detailed in \cite{murray_dermott99} is 
%
\begin{equation} \label{eq:eq1}
v_{ {\rm{rad}}, p }\,\approx\,\frac{m_{s} }{m_{s}+m_{p}} n  a  \sin{i} \left [ \sin(M+\overline\omega) +e \sin(2M+\overline\omega) \right ],
\end{equation}
%
with all quantities defined in Table\,1.\\


An eclipse or occultation of a part of the rotating planetary surface causes a time-dependant asymmetry in the absorption/emission line profiles. These asymmetries result in an apparent shift of the central spectral line positions when the lines are unresolved. 

\begin{figure}[t]
\centering
\includegraphics[scale=0.725,keepaspectratio]{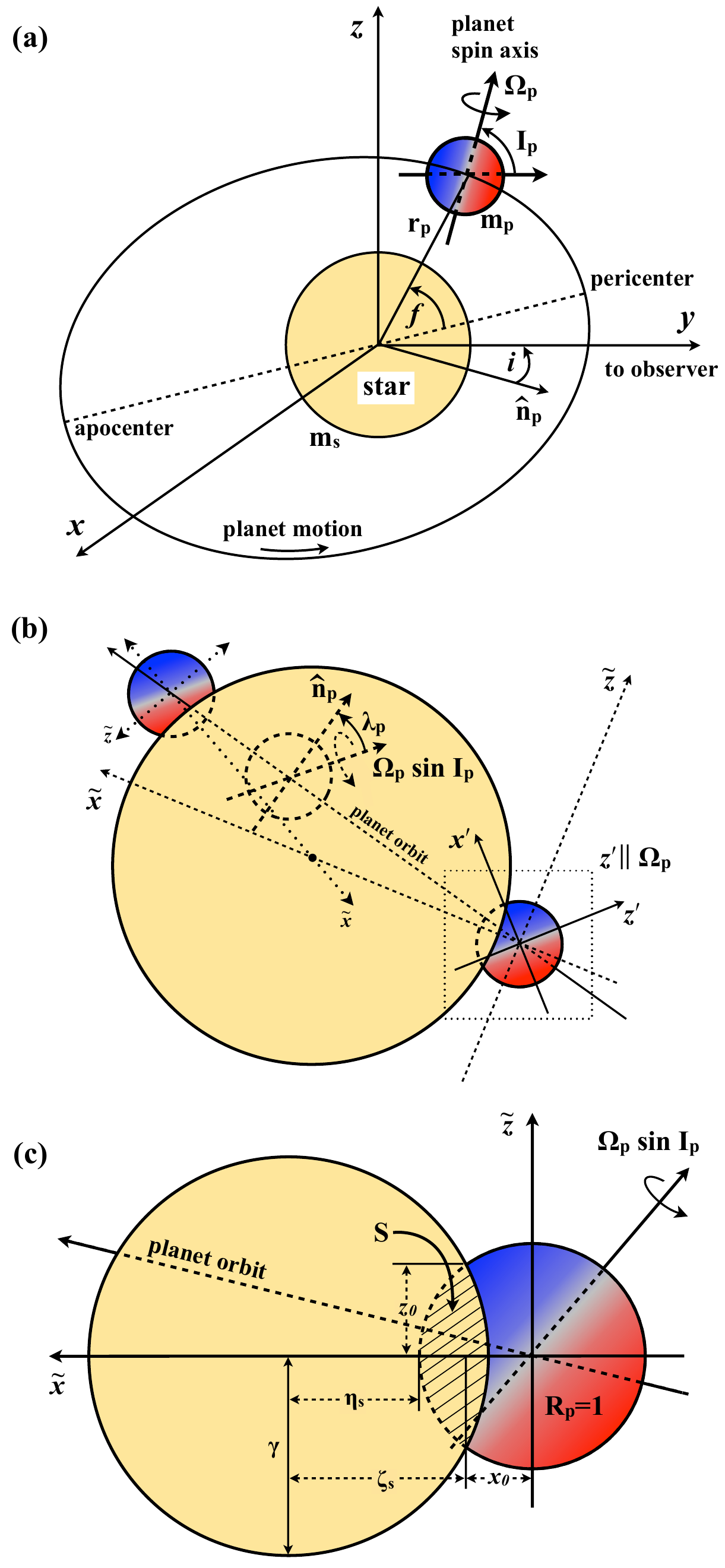}
\caption{(a): Schematic illustration of the planetary orbit plane, spin axis, and the observer's line of sight; (b): Planet secondary eclipse ingress and egress phases and rotation axis; (c): A zoom of the planet and star configuration at ingress in the new coordinates. (see Table\,\ref{table:notation} for symbol definitions).}
\label{fig:FIG1}
\end{figure}

To describe the radial velocity anomaly caused by the planet rotation, similar to \cite{ohta05}, we initially set the coordinate system at the star centre and its $y$-axis to coincide with the observer's line of sight (Fig.\,\ref{fig:FIG1}\,a). The planet position is described with the coordinates $(x_p, z_p)$, corresponding to the orbit plane position and the planet impact parameter.\,


For simplicity of the mathematical description of the problem, we choose a reference system ($x^{\prime}, z^{\prime}$) centred on the planet and rotated such that the $z^{\prime}$-axis is parallel to the rotation axis of the planet (i.e. parallel to $\Omega_p$, see Fig.\,\ref{fig:FIG1},\,b) and the rotation axis lies in the $y^\prime$-$z^\prime$ plane. We define an angle $\lambda_p$ between the sky-projected rotational angular velocity and the normal unit vector of the planet orbit, $\hat{n}_p$, see Fig.\,\ref{fig:FIG1},\,b. This differs from the definition of \cite{ohta05}, who assume $\lambda$ to be the angle between the sky-projected stellar rotation axis $\Omega_s$ and the normal vector of the planetary orbit $\hat{n}_p$.



%
%

%
%
%

\begin{figure}
\centering
\includegraphics[scale=0.51,keepaspectratio]{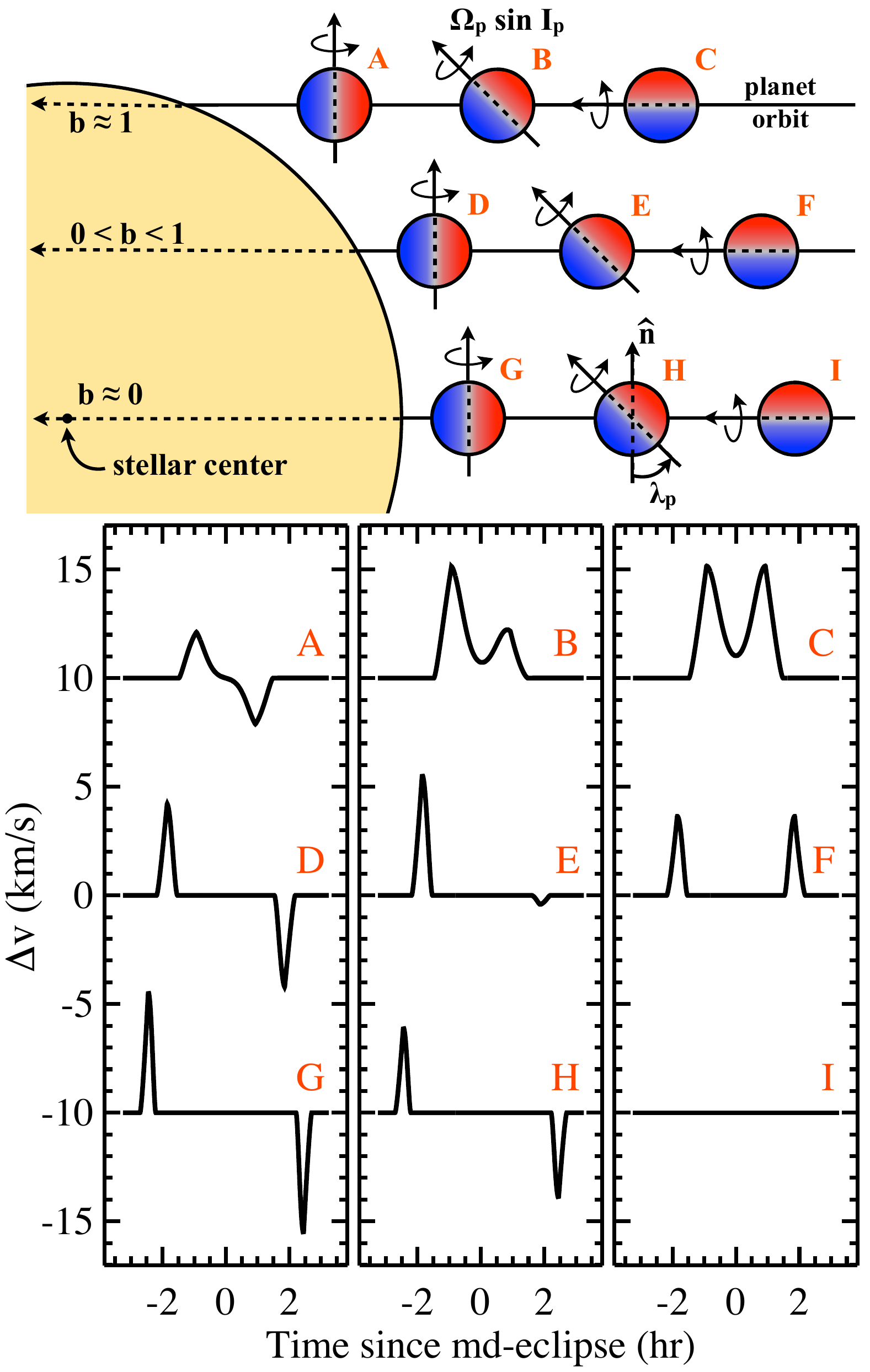}
\caption{ Illustration of planet radial velocity curve anomaly due to \textit{RMse} effect for nine (representative) prograde spin-orbital alignments. The curves are plotted with a constant 10\,km\,s$^{-1}$ offset for clarity. Cases A to I correspond to the top planet-star configurations. The curves flip and invert as $\lambda_p$ increases to $360^{\circ}$ and when b\,$<0$.}
\label{fig:FIG3}
\end{figure}

%
In all calculations we ignore differential rotation of the planet surface as well as motions associated with atmospheric dynamics. A point on the surface of the rotating planet with coordinates $(x^\prime, z^\prime)$ will move with a velocity $v_p$ given by 
%
\begin{equation} \label{eq:eq2p0}
v_{ p } = \Omega_p x^\prime \sin{I_p},
\end{equation}
%
where $\Omega_{p} $ is the angular velocity of the planet. The associated radiation will exhibit a Doppler shift defined as
%
\begin{equation} \label{eq:eq2p1}
\frac{\Delta \nu}{\nu} = \frac{ \Omega_p x^\prime \sin{I_p} }{c},
\end{equation}
%
with respect to the observer along the $y^\prime$-axis (i.e. the line of sight). We refer the reader to Sec.\,3, Equations (14-19) in \cite{ohta05} for a derivation of the radial velocity profile for a star and adopt their expression (20) rewritten for the planet:
%
\begin{equation} \label{eq:eq2}
\Delta v_{ p } = - \Omega_{p} \sin{I_{p}} \frac{ \iint{x^{\prime}I(x^{\prime},z^{\prime})dx^{\prime}dz^{\prime}}}{ \iint{I(x^{\prime},z^{\prime})dx^{\prime}dz^{\prime} }},
\end{equation}
%
Equation~\ref{eq:eq2} relates the radial velocity change $\Delta v_p$ and the line intensity $I(x^{\prime},z^{\prime})$. Fig.\,\ref{fig:FIG3} illustrates the different cases of the RM effect during secondary eclipse.

We evaluate the integrals assuming uniform model of the planet surface intensity $I(x^{\prime},z^{\prime})$. We ignore the role of planet limb-darkening as our goal is to estimate the first order rotational effect, and leave inclusion of the limb-darkening for future investigations. We also consider the star to be completely optically thick.
%
%
%
%
%
%

%
%
%
At ingress and egress the position of the stellar disc satisfies the relation $R_s-R_p<(x_s^{\prime2}+z^{\prime2}_s)^{1/2} < R_s+R_p$. In order to simplify the computational task we rotate the coordinates in a time-dependent manner so that the stellar centre is always located along the new $\tilde{x}$-axis, as in \cite{ohta05}, see Fig.\,\ref{fig:FIG1}\,b:
%
\begin{equation} \label{eq:eq6}
\begin{pmatrix}
  \tilde{x}  \\
  \tilde{z}  
 \end{pmatrix}
 =
\frac{1}{R_p\sqrt{x^{\prime2}_s+z^{\prime2}_s}}
\begin{pmatrix}
  x^{\prime}_{s}  & z^{\prime}_{s}\\
  -z^{\prime}_{s} & x^{\prime}_{s}
 \end{pmatrix}
\begin{pmatrix}
  x^{\prime}\\
  z^{\prime}
 \end{pmatrix}.
\end{equation}
%
The position of the star is given by
%
%
\begin{equation} \label{eq:eq7}
\begin{pmatrix}
  \tilde{x}_s  \\
  \tilde{z}_s  
 \end{pmatrix}
 =
\begin{pmatrix}
  1+\eta_{s}  \\
  0
 \end{pmatrix},
\end{equation}
%
where $\eta$ is defined as
%
\begin{equation} \label{eq:eq8}
\eta_{s} = \sqrt{\frac{x_s^{\prime2}+z^{\prime2}_s}{R_p^2}} -1.
\end{equation}
%
The intensity on the uniform planet surface at ($\tilde{x}, \tilde{z}$) is given by
%
\begin{equation} \label{eq:eq9}
 I(\tilde{x},\tilde{z})=
\begin{cases}  I_0, &\tilde{x}^2 + \tilde{z}^2 \leq 1~\text{and}~(\tilde{x}-1-\eta_s)^2+\tilde{z}^2 \geq \gamma^2,
\\
0 &\text{otherwise},
\end{cases}
\end{equation}
%
where $\gamma=R_s/R_p$, $R_p=1$. The moments of the intensity then are
%
\begin{equation} \label{eq:eq10}
 \iint I(x^{\prime}, z^{\prime})~\mathrm{d}{x^{\prime}}\,\mathrm{d}{z^{\prime}} = R_p^2 \bigg[ \pi I_0 - \iint_S I(\tilde{x},\tilde{z})~\mathrm{d}{\tilde{x}}\,\mathrm{d}{\tilde{z}}\bigg],
\end{equation}
%
and
%
\begin{equation} \label{eq:eq11}
 \iint x^{\prime}I(x^{\prime}, z^{\prime})~\mathrm{d}{x^{\prime}}\,\mathrm{d}{z^{\prime}} = -\frac{R_p^2}{1+\eta_s} \iint_S (x^{\prime}_s \tilde{x} - z^{\prime}_s \tilde{z})I(\tilde{x},\tilde{z})~\mathrm{d}{\tilde{x}}\,\mathrm{d}{\tilde{z}}.
\end{equation}
%
The range of the integrals is denoted with $S$ and is defined as the star-planet overlapping region (shaded area) in Fig.\,\ref{fig:FIG1}\,c:
%
\begin{equation} \label{eq:eq12}
 \iint_S \mathrm{d}{\tilde{x}}\,\mathrm{d}{\tilde{z}} = \int_{x_0}^{1} \mathrm{d}{\tilde{x}}    \int_{-\sqrt{1-\tilde{x}^2}}^{\sqrt{1-\tilde{x}^2}} \mathrm{d}{\tilde{z}} + \int_{\tilde{x}_s-\gamma}^{x_0} \mathrm{d}{\tilde{x}}    \int_{ -\sqrt{ \gamma^2-(\tilde{x}-\tilde{x}_s)^2}}^{ \sqrt{ \gamma^2-(\tilde{x}-\tilde{x}_s)^2}  } \mathrm{d}{\tilde{z}}  ,
\end{equation}
%
The star and planet circles intersect at ($x_0, \pm z_0$), where
%
\begin{equation} \label{eq:eq13}
x_{0} = 1- \frac{\gamma^2-\eta_s^2}{2(1+\eta_s)},
\end{equation}
%
%
\begin{equation} \label{eq:eq14}
z_{0} = \sqrt{1-x_0^2} = \frac{ \sqrt{(\gamma^2-\eta_s^2) [(\eta_s+2)^2-\gamma^2]} }{ 2(1+\eta_s) }.
\end{equation}
%
These calculations are equivalent to the calculations presented in \cite{ohta05} and to simplify the final result we introduce
%
\begin{equation} \label{eq:eq15}
\zeta = 1+\eta_s-x_0 = \frac{ 2\eta_s+\gamma^2+\eta_s^2 }{ 2(1+\eta_s) }.
\end{equation}
%
Then equations \ref{eq:eq10} and \ref{eq:eq11} are analytically integrated as
%
\begin{equation} \label{eq:eq16}
 \iint_S I(\tilde{x},\tilde{z}) \mathrm{d}{\tilde{x}}\,\mathrm{d}{\tilde{z}} = I_0 \bigg[\sin^{-1}{z_0} - (1+\eta_s)z_0
+\gamma^2\cos^{-1}{\zeta/\gamma}\bigg],
\end{equation}
%
and
%
\begin{equation} \label{eq:eq17}
\begin{split} 
 \iint_S (\tilde{x}x^{\prime}_s-\tilde{z}z^{\prime}_s)I(\tilde{x},\tilde{z})\mathrm{d}{\tilde{x}}\,\mathrm{d}{\tilde{z}} = I_0x^{\prime}_s(1+\eta_s)\bigg[-z_0\zeta \\
 +\gamma^2\cos^{-1}{\zeta/\gamma} \bigg].
\end{split} 
\end{equation}
%
Finally combining these two results we find the radial velocity anomaly of the planet (during ingress or egress) as a function of the star position ($x^{\prime}_{s}$):
%
\begin{equation} \label{eq:eq18}
\Delta v_{p} = \Omega_{p} x^{\prime}_{s} \sin{I_{p}} \frac{-\rm{z}_{0}\zeta + \gamma^2 \cos^{-1}(\zeta/\gamma) }{ \pi - \sin^{-1}{\rm{z}_{0}} + (1+\eta_{s})\rm{z}_{0} - \gamma^{2}\cos^{-1}(\zeta/\gamma)},
\end{equation}
The final result is equivalent to the result of \cite{ohta05} which is expected given the identical overlapping area between the planet and star circles.
%

%
%
%
%
%
%
%
%
%
%
\section{ \textit{RMse} effect amplitude and shape }\label{sec:reas}

The shape and amplitude of the \textit{RMse} effect are illustrated in Fig.\,\ref{fig:FIG3} for nine representative configurations of the planet spin-orbital alignments (i.e. impact parameter $b = \frac{a}{R_{\ast}} \cos{i}$ and $\lambda_p$), corresponding to prograde rotation. In case of retrograde rotation (i.\,e. $\lambda_p\geq90^{\circ}$) the radial velocity curves are inverted. The \textit{RMse} computation in Fig.\,\ref{fig:FIG3} assumes a Jupiter-like planet, with $v\sin{I_p}$ at the equator of $12.6$\,km\,s$^{-1}$, on a $20$\,day orbit around a Sun-like star. We choose a period rather longer than the currently typical hot-Jupiter orbital periods because close-in exoplanets are expected to rapidly synchronise their rotation with the orbital period due to tides raised by the star on the planet. The time to spin-down the planet rotation is given by 
\begin{equation} \label{eq:eq19}
\tau_{syn}\approx  Q_p \Big(\frac{R_p^3}{GM_p}\Big) (\omega-\omega_s) \Big(\frac{M_p}{M_{\ast}}\Big)^2 \Big(\frac{a_p}{R_p}\Big)^6,
\end{equation}
%
where $Q_p$, $R_p$, $M_p$, $\omega$, $\omega_s$, $M_{\ast}$ and $a_p$ are the planet's tidal dissipation factor, radius, mass, rotational angular velocity, and synchronous (or orbital) angular velocity \citep{Goldreich66, guillot96}. Assuming $Q_p\sim10^5$ \citep{correia10} and orbital periods of 4 and 20\,d, we find $\tau_{syn}\sim10^6$ and $10^9$\,yr, respectively implying that a $\sim20$\,d orbit required to have spin-down times comparable to the ages of the currently observed transiting hot Jupiters.

Our choice for the value of $v\sin{I_p}$ is driven by the correlation of the equatorial rotational velocities of solar system planets and their masses, suggesting that more massive planets rotate faster \citep{hughes03}. Currently a rotational rate has been constrained only for the massive and young planet $\beta$\,Pic\,b ($v\sin{I_p}\sim25$ \,km\,s${-1}$), which is in line with the spin velocity-mass relation of the solar system \citep{snellen14}.

The maximum amplitude of the \textit{RMse} effect is expected for central eclipses (i.\,e.~$\text{b}=0$) and $\lambda_p=0~\text{or}~180^{\circ}$. In the case considered in Fig.\,\ref{fig:FIG3}, the amplitude is ${\pm\sim6\,\text{km\,s}^{-1}}$. Assuming the $v\sin{I_p}$ of $\beta$\,Pic\,b we find rotation effect with an amplitude of $\sim11~\text{km\,s}^{-1}$. No effect is expected in cases where equal portions red- and blue-shifted portions of the planet are eclipsed (i.e. $\text{b}=0$ and $\lambda_p=90^{\circ}~\text{or}~180^{\circ}$). This is also the case where the planet rotation axis is normal to the plane of the sky. 

It should be pointed out that in all cases that include $b=0$, $v\sin{I_p}$ is degenerate with $\lambda_p$, i.e. various combinations of these two parameters can produce the same radial velocity amplitude. Importantly, it is the amplitude difference of the ingress and egress \textit{RMse}, their signs and the shape of the radial velocity curve that constrain the planet obliquity and break this degeneracy. 

\begin{figure}
\centering
\includegraphics[scale=0.83,keepaspectratio]{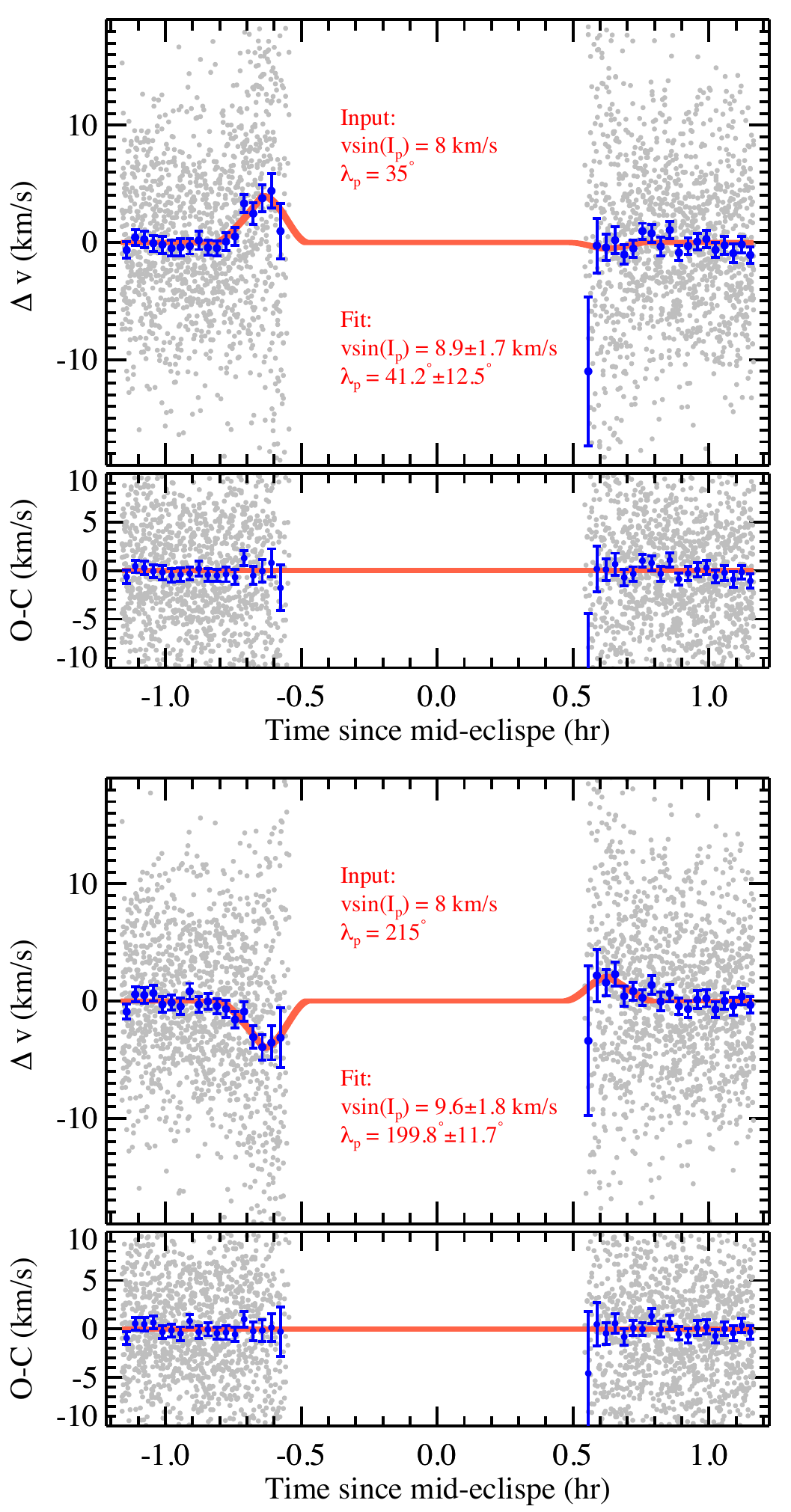}
\caption{Mock data (grey dots) illustrating the \textit{RMse} effect, best-fit radial velocity curve (red lines) and radial velocity residuals. Upper and lower pair panels illustrate the cases of a WASP-19 system, assuming tidally synchronised prograde and a retrograde rotating planet, respectively. The blue symbols indicate the binned radial velocity curves by 2\,min.}
\label{fig:simulation_result}%
\end{figure}


\section{ An observational perspective for \textit{RMse} }\label{sec:obs_persp}

We produce mock data to illustrate the planetary \textit{RMse} effect assuming an 8 and 40\,m class telescopes, equipped with a near-infrared, high spectral resolution (i.e. $\text{R}\sim100,000$) spectrometer. We assume that the planet orbit is constrained with enough precision to be subtracted prior to the search for the \textit{RMse} effect. We calibrate our simulation adopting a precision of $\sim5$\,km\,s$^{-1}$ as achieved by \cite{Birkby13} at $3.2\,\mu$m for water detection in the day side atmosphere of HD\,189733b. We scale this precision by the square root of the number of collected spectra (i.e. 48), the ratio of the employed spectral coverage (assuming a hypothetical high-resolution spectrometer with a wavelength coverage of 400\,nm) and a square root of the ratio of the target planet-to-star flux ratio and the planet-to-star flux ratio of HD\,189733b, i.e. $1.3\times 10^{-3}$ from \cite{Birkby13}. This factor plays a critical role as it accounts for the strength of the planet signal, which is determined by the planet to star flux contrast. The contrast decreased from $\sim10^{-3}$ to $\sim10^{-6}$ when increasing the planet orbital periods from $\sim0.5$\,day to $20$\,day. We also factor in the sampling rate with the telescope diameter and the brightness of the target host star. An account for the change of the planet flux during a secondary eclipse is also incorporated by factoring the radial velocity uncertainties  by square root of the flux of a secondary eclipse \cite{mandel02} model with unity out-of eclipse baseline and zero in-eclipse flux. 

Fig.\,\ref{fig:simulation_result} shows mock data generated assuming a 40\,m-class telescope and a hypothetical planet host star of brightness $\text{K}=5.5$\,mag (i.e. similar to the currently brightest transiting hot-Jupiter host star HD\,189733). Fig.\,\ref{fig:simulation_result} shows the results for a planet with the physical properties of WASP-19\,b, which is considered to be tidally locked. We assumed a prograde ($\lambda_p = 35^{\circ}$) and a retrograde ($\lambda_p = 215^{\circ}$) rotation with $v\,\sin{I_p}=8$\,km\,s$^{-1}$. We perform a \textit{RMse} fit to the mock data utilising a Levenberg-Marquardt least-squares minimization and estimate $\lambda_p$ and $v\sin(I_p)$, \citep{Markwardt09}. We find the planet signal to be well-detected ($\sim5$-$\sigma$) when combining nine secondary eclipses.

We also explored the potential of an 40\,m-class telescope to detect the \textit{RMse} effect in the currently brightest transiting exoplanet host-star HD\,189733\,b. Assuming a synchronised planet rotation (i.e. \textit{RMse} amplitude of $\sim1$\,km\,s$^{-1}$) and $\lambda_p > 0^{\circ}$ we find that $\sim50$ secondary eclipses need to be co-added to detect the spin of HD\,189733\,b. 

Finally, we investigated the potential of an 8\,m-class telescope to detect the \textit{RMse} effect in K=5.5 host star with a WASP-19b-like planet and synchronised rotation. We find a number of $\sim50$ secondary eclipses will have to be added in order to detect the \textit{RMse} effect at $\sim3$-$\sigma$ confidence.

\section{Discussion and conclusions}\label{sec:discussionsec}

We have discussed the Doppler signature of a spinning exoplanet during secondary eclipse  and derived a relation for the radial velocity as a function of the planet's rotational rate and axial tilt (Eq.\,\ref{eq:eq18}). The \textit{RMse} effect complements the tool-box offered by transiting exoplanets providing a proxy to exoplanet spins. We showed that the formalism of the \textit{RMse} is equivalent to the formalism developed by \cite{ohta05} to describe the RM effect caused by a planet transiting its star just from a different perspective. We note that \cite{hirano11} derived an improved accuracy radial velocity anomaly curves for the RM effect during primary transits. However, the solution of \cite{ohta05} provides a conservative estimate of the radial velocity amplitude and is precise enough to illustrate the \textit{RMse} effect.

We acknowledge that our assumption of a rigid rotating non-limb darkened planetary surface (i.e. ignoring differential rotation and atmospheric dynamics) is a crude approximation, but it illustrates the expected \textit{RMse} radial velocity anomaly in the simplest case. The radial velocity originating from the \textit{RMse} effect is expected to be degenerate with atmospheric dynamics. Exoplanets are expected to have atmospheric circulation with increasing dynamics at shorter orbital distances \citep{Kataria15}. Therefore it is expected that the radial velocity curves of most close-in exoplanets may significantly differ from that assumed here, with super-rotating winds expected to enhance the radial velocity signals by factor of a few. The \textit{RMse} effect presents the opportunity to probe atmospheric winds with altitude, when wide coverage high dispersion spectroscopy becomes capable of providing per-point precisions in the $\sim$m\,s$^1$ domain. In such cases it could be possible to measure longitudinally-integrated radial velocities and map the surface of a planet during the secondary eclipse ingress/egress phases, providing constraints on the hot-spot position. We postpone the derivation of the \textit{RMse} effect in these cases for future theoretical global circulation model investigations and identify the top three planets expected to exhibit the strongest \textit{RMse} effect assuming synchronised rotation. Applying Eq.\,\ref{eq:eq18} (assuming $\lambda_p=0^{\circ}$) we find values of $\Delta v_p\sim3.7,~3.4,~\text{and}~3$\,km\,s$^{-1}$ for targets WASP-103\,b, WASP-12\,b and WASP-19\,b, respectively. 

\cite{kawahara12} demonstrated the rotational signature of an exoplanet in the planet radial velocity. The \textit{RMse} effect discussed in this study is of similar nature. However the geometric shadow of the star rather than the changing planet phase is the factor removing velocity components from the visible planet surface. Importantly, because the \textit{RMse} effect occurs only during ingress and egress, its detection requires significantly less observing time compared to the case in \cite{kawahara12}. \textit{RMse} therefore could be much-well suited for planets with longer orbital periods.

We also demonstrated that to be put into practice the \textit{RMse} effect requires exoplanet host stars brighter than the currently known exoplanet hosts, i.e. brighter than $\text{K}\sim6$\,mag and large aperture telescopes (i\,e.\,$\sim$40\,m). The strongest \textit{RMse} effect for evolved stars is expected for non-synchronized rotation, i.e. planets on longer than 20 day orbits, with a typical $v\sin{I_p}~12.5$\,km\,s$^{-1}$ or more. However, such planets exhibit lower temperatures (i.e.\,$<700$\,K), compared to the typical hot Jupiters giving a small planet-to-star flux contrasts (i.e. $\sim 10^{-6}$ or lower at $\sim3\,\mu$m). The detection of the \textit{RMse} effect in such cases would be an extremely challenging if not impossible task, even for the upcoming 40\,m class telescopes. 

An interesting case could be young planets which still have not been synchronised. Observations using high-dispersion, near-infrared spectroscopy have currently constrained the rotational rate of the only one such case ($\beta$\,Pic\,b) from rotationally broadened absorption lines \citep{snellen14}. An important opportunity could exist if the orbit orientation of this planet allows transits and secondary eclipses, because the large brightness of the host star (i.\,e. $\text{K}\sim3.5$\,mag) and the fast spin of the planet (i.\,e. $v\sin{I_p}\sim27$\,km\,s$^{-1}$) would produce a strong and detectable \textit{RMse} signal. Another interesting opportunity could be offered by transiting brown dwarfs with orbital periods larger than 10-20 days. Such systems assume high flux contrasts between the star and the brown dwarf in the near-infrared which in case of a bright host star (i.\,e. $\text{K}\sim5$\,mag) could provide an opportunity to probe the spins and latitudinal radial velocity maps of these objects.

In the near future, the Next Generation of Transit Surveys (NGTS), the Transiting Survey Satellite (TESS) and the Planetary Transits and Oscillations of stars (PLATO) projects are expected to significantly expand the sample of known planets hosted by bright stars (i.e. brighter than $\text{K}\sim9.5$\,mag) and hence to provide more targets suitable for detection of the \textit{RMse} effect. 

The hypothetical radial velocity data in Section\,\ref{sec:obs_persp} assumes retrieval of the planet signal via cross-correlation with a model planet spectrum similar to \cite{Snellen10, Brogi12, Birkby13, deKok13, Schwarz15}. Although currently the cross-correlation technique has not been demonstrated to provide per spectrum radial velocity measurements, in their Fig.\,1, \cite{Snellen10} presented results for the planet geocentric radial velocity as a function of the planet orbital phase at a high significance ($\sim3$-$\sigma$). It is expected that large-aperture near-future telescopes (i.e.\,30-40\,m) could provide even higher significance signals and planet radial velocity measurements from multiple spectra. In addition, cloud-free exoplanet atmospheres are expected to result in stronger cross-correlation signals compared to hazy or cloudy atmospheres for the same stellar and planetary physical properties. Cloud-free atmospheres are expected to have well-pronounced lines, i.e. non-muted absorption features which are expected for hazy atmospheres.

 Stellar activity and pulsations are known to produce radial velocity variations that could in some cases mimic those induced by the orbital motion of exoplanets. This may lead to misinterpretations of radial velocity variations, especially when those variations have periods less or equal to the star rotational periods \citep{Lagrange10}. Therefore, it is important to consider whether stellar activity, for instance the distortion to the stellar spectral lines induced by starspots on a rotating stellar surface, could induce residual signal misinterpreted as exoplanet signal. This is especially pertinent for co-added observations obtained over several nights.

 
 A typical \textit{RMse} observation would last for $\sim0.8$\,h on each of the ingress and egress phases of a WASP-19b-like planet (see Fig.\,\ref{fig:simulation_result}). This time interval is significantly shorter than the rotation periods of the typical planet host stars ranging from 10 to 40 days \citep{Paz15}. Furthermore, starspots are expected to produce sine-wave radial velocity variations (i.e. different from the \textit{RMse} shape) with amplitudes of a few hundred m\,s$^{-1}$ in the optical and are expected to reduce by factor of at least a few at near infrared wavelengths, where the contrast between the photosphere and cool star spots (e.g. $\Delta\,T\sim550$K) is significantly reduced. 

Thus, if present, residual stellar signal caused by starspots would be a factor of a few tens smaller than the the expected 1-2\,km\,s$^{-1}$ amplitude of the \textit{RMse} effect for the closest synchronized planets and even more for non-synchronized planets on longer than 20 day period orbits. Only in the case of most evolved planets (i.e. typical hot Jupiters), if synchronized, with orbital/rotational periods from 4-5 days to $< 20$ day could a residual introduce larger radial velocity scatter when multiple observations are combined. However, such planets are expected to have \textit{RMse} amplitudes much smaller than 1\,km\,s$^{-1}$ and would be difficult targets by definition.

	In the near infrared the planet spectrum is dominated by thermal radiation and is not expected to contain information from distortions caused by starspots. In case that the planet spectrum also contains a non-negligible component from a reflected star spectrum then the planet spectral lines will have distortions caused by the star spots. This would make the planet spectrum different and more difficult to cross correlate with model spectra. However, such an effect is not expected to mimic the \textit{RMse}.

In conclusion, we have demonstrated the \textit{RMse} effect and its potential to constrain planet rotational rates and sky projected spin-orbital alignments. We derived the radial velocity curve caused by the RM anomaly and estimated the amplitude of the effect for a hot Jupiter. Finally we discussed the prospects for detecting the effect and constraining planet spins and axial tilts from the current and upcoming instrumentation.

\acknowledgments
We are grateful to D. Sing, T. Evans, T. Kataria, M. Browning, H. Wakeford and A. Nikolova for valuable and pleasant discussions that helped to improve the manuscript. We are grateful to the anonymous Referee for their valuable comments and suggestions for improving the manuscript. NN acknowledges support from an STFC consolidated grant. FSM acknowledges the University of Exeter and the European Union's Horizon 2020 research and innovation program under ERC starting grant agreement No.\,337705\,(CHASM).

\end{document}